\documentclass[letter,twocolumn]{jpsj2}
\usepackage{graphicx}
\usepackage{color}

\title{Evolution of the spin resonance in CeCoIn$_5$ under magnetic field}

\author{%
J. Panarin, S. Raymond\thanks{Corresponding author, E-mail: raymond@ill.fr}, G. Lapertot and J. Flouquet
}

\inst{%
CEA-Grenoble, INAC / SPSMS, 38054 Grenoble, France
}

\recdate{\today}

\abst{The evolution of the magnetic excitation spectrum of the unconventional superconductor CeCoIn$_{5}$ was studied by inelastic neutron scattering  as a function of magnetic field applied in the [1, -1, 0] direction. The spin resonance characteristic of the superconducting states was observed up to about $H_{c2}$/2 : it is still centered at (1/2, 1/2, 1/2) when the magnetic field increases while its characteristic energy decreases and its lineshape substantially broadens.}

\kword{%
unconventional superconductivity, heavy fermion, inelastic neutron scattering.
}

\begin{document}
\maketitle
The interplay between magnetism and superconductivity is one of the most studied topic in the physics of strongly correlated electron systems. To this aim, the family of heavy fermion (HF) compounds CeMIn$_5$ (M = Co, Rh, Ir), the so-called 1-1-5 compounds, is a fabulous playground since the chemical doping, the application of pressure and magnetic field leads to the possibility to tune the N\'eel temperature ($T_{N}$) and the superconducting transition temperature ($T_{c}$) to different levels. Compare to other HF systems, this tuning is realized up to $T_{N}$=$T_{c}$ allowing fine studies  of the competition between magnetism and superconductivity \cite{revue}. Among the 1-1-5 family, CeCoIn$_{5}$ is the HF compound with the highest superconducting transition temperature of 2.3 K and a moderate Sommerfeld coefficient $\gamma$=290 mJ/molK$^2$ just above $T_{c}$. However the huge value of the specific heat jump at $T_{c}$ points toward a strong coupling situation. In the absence of superconductivity, the Sommerfeld coefficient will be $\gamma_{0}$ $\approx$ 1000 mJ/molK$^2$. As concern the low energy magnetic excitations measured by Inelastic Neutron Scattering (INS), a quasielastic signal is measured above $T_{c}$ with a relaxation rate of 0.3 meV that is consistent with the huge value of $\gamma_{0}$.
Below $T_{c}$, the formation of Cooper pairs with coherence factor characteristic of the $d_{x^2-y^2}$ pairing is manifested by a switch from a quasielastic to a sharp inelastic peak that appears for an energy of 0.6 meV ($\approx$ 3k$_{B}$$T_{c}$)  at the antiferromagnetic position (1/2, 1/2, 1/2) \cite{Stock}. This kind of low energy excitation is also observed in the HF compounds UPd$_{2}$Al$_{3}$ \cite{Metoki} and CeCu$_{2}$Si$_{2}$ \cite{Stockert}). This mode is observed at the hot spots of the Fermi surface where the magnetic fluctuations are peaked. It is named  resonance in analogy with the studies performed in the high-T$_{c}$ Superconductors (HTSC) \cite{Sidis}.

The behavior of CeCoIn$_{5}$ under magnetic field opens new views on magnetism and superconductivity. It is indeed believed that a field induced magnetic quantum critical point occurs in this compound near the superconducting upper critical field $H_{c2}$ and that the superconducting gap precludes the formation of antiferromagnetism when $T_{c}$ overpasses what will be $T_{N}$ in the absence of superconductivity\cite{Ronnig}. Moreover, for a magnetic field applied in the basal plane of the tetragonal structure, a new phase (named here LTHF for Low Temperature High Field) appears in a narrow range of temperature and magnetic field below 350 mK and above 10.5 T, the upper critical field being 11.6 T in this direction. This phase is a prime candidate for the realization of a FFLO (Fulde, Ferrel, Larkin, Ovchinnikov) ground state \cite{Revue2}. Interestingly field induced long range incommensurate order is reported in this phase by neutron diffraction, and solely in this phase \cite{Kenzelmann}. The propagation vector of the LTHF phase is (0.44, 0.44, 1/2). In this letter we report the evolution of the so-called resonance under magnetic field applied along the [1, -1, 0] direction with focus on the field range where the vortex density remains smaller than the superconducting one ($H <  H_{c2}/2$).

The experiments were performed on the cold neutron triple-axis spectrometers IN14 and IN12 at ILL, Grenoble. The incident beam was provided by a vertically focusing pyrolytic graphite (PG) monochromator. A liquid-nitrogen-cooled Be filter was placed just before the sample in order to cut down the higher-order contamination of  neutrons.  A horizontally focusing PG analyzer was used on IN12 and a double focusing analyzer was used on IN14. Measurements were performed with a fixed final wave vector k$_{f}$ of 1.3 \AA$^{-1}$.  The collimations were 60'-open-open. The energy resolution determined by the Full Width at Half-Maximum (FWHM) of the incoherent signal was 0.1 meV on IN12 and 0.15 on IN14. However the incoherent tail extends up to 0.3 meV for both setups. The sample consists of about 40 single crystals of CeCoIn$_{5}$ co-aligned on two aluminium plates and fixed with Fomblin oil. The mosaic spread of such an assembly is of one degree as measured on a rocking curve of the (1, 1, 1) Bragg reflection. This assembly was put in a dilution insert inside the 15 T and 12 T vertical field magnets with the field applied along the [1, -1, 0] direction and the scattering plane was therefore ($h$, $h$, $l$).
\begin{figure}[t!]
\centering
\includegraphics[width=8cm]{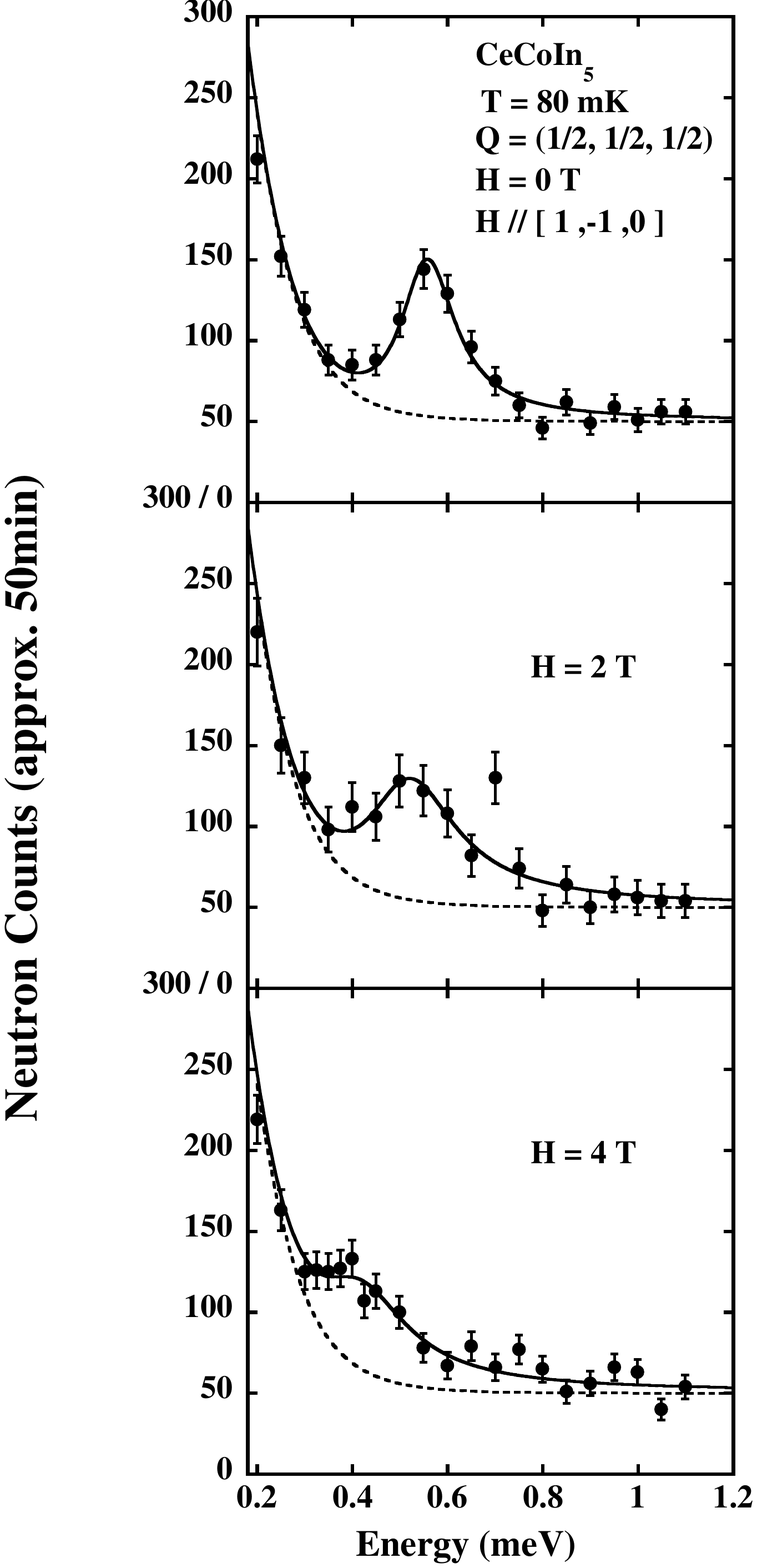}
\caption{Excitation spectrum measured at $\bf{Q}$=(1/2, 1/2, 1/2) for $H$ = 0, 2 and 4 T and $T$=80 mK. The solid lines are Lorentzian fits as described in the text. The dashed line indicates the background measured at $\bf{Q}$=(0.2, 0.2, 1/2) at 4 T and 80 mK.}
\end{figure}

Representative spectra measured at $\bf{Q}$=(1/2, 1/2, 1/2) are shown in Figure 1 for $H$ = 0, 2 and 4 T at 80 mK. The data at zero field are consistent with the one reported by Stock et al. \cite{Stock} except for a slight difference in the peak position that is found here at 0.55 meV. The position of the resonance shifts to lower energy when the magnetic field increases while the spectra broadens. The dashed line is a fit to the background spectrum measured at $\bf{Q}$ = (0.2, 0.2 ,0.5) at $H$ = 4 T and 80 mK. Constant energy scans were performed at 0 and 4 T and at 80 mK in order to locate the peak position in $\bf{Q}$ space under magnetic field. Such spectra (shown in Figure 2) were measured at the maximum of the resonance peak i.e. at 0.55 meV for 0 T and 0.4 meV for 4 T for both [1, 1, 0] and [0, 0, 1] directions around (1/2, 1/2, 1/2). It is found that the peak is also located at (1/2, 1/2, 1/2) at 4 T. The constant energy scans were analyzed by Gaussian lineshape and the correlations lengths obtained from the inverse of the gaussian half width at half maximum are $\xi_{a}$=12.2 (7) $\AA$ and $\xi_{c}$=10.0 (5) $\AA$. Compared to the work of Stock\cite{Stock}, we found more isotropic correlations. Within our precision, the correlation lengths do not change under magnetic field. The superconducting coherence length is of about 30 $\AA$ \cite{Knebel2}, that is three times the magnetic correlation length. The inter-vortex distance is much greater of about 200 $\AA$ at 5 T. Note that the background of the constant 
energy scans at 0 and 4 T is consistent with the background of the constant $\bf{Q}$ scan measured at (0.2, 0.2 ,0.5) at $H$ = 4 T. The constant $\bf{Q}$ spectra were analyzed by a Lorentzian lineshape with an inelasticity $\Delta_{\bf{Q}}$. The measured neutron intensity subtracted from the background is proportional to the scattering function $S(\textbf{Q},E)$. We analyzed our data using :
\begin{equation}
S(\textbf{Q},E)=\frac{1}{1-exp(-E/k_{B}T)}\times\frac{\chi'_{\textbf{Q}}\Gamma_{\textbf{Q}}E}{(E-\Delta_{\textbf{Q}})^2+\Gamma_{\textbf{Q}}^2}
\end{equation}
where $\Gamma_{\textbf{Q}}$ is the relaxation rate and $\chi'_{\textbf{Q}}$ is the susceptibility at the wavector $\bf{Q}$. The energy spectra were taken only for a single $\bf{Q}$=(1/2, 1/2, 1/2), we therefore drop the suffix $\textbf{Q}$ in the following. Since the data are not normalized into absolute units, $\chi'$ is given in arbitrary units. Figure 3 shows the field evolution of $\chi'$ and of the linewidth $\Gamma$ and Figure 4 shows the field  dependence of the energy gap $\Delta$. Within the error bars, $\chi$' remains constant under magnetic field while the relaxation rate increases being from 0.07(1) meV at zero field to 0.15(3) meV at $H$ = 5 T. The energy gap decreases almost linearly.

\begin{figure}[t!]
\vspace{-0.7cm}
\centering
\includegraphics[width=8cm]{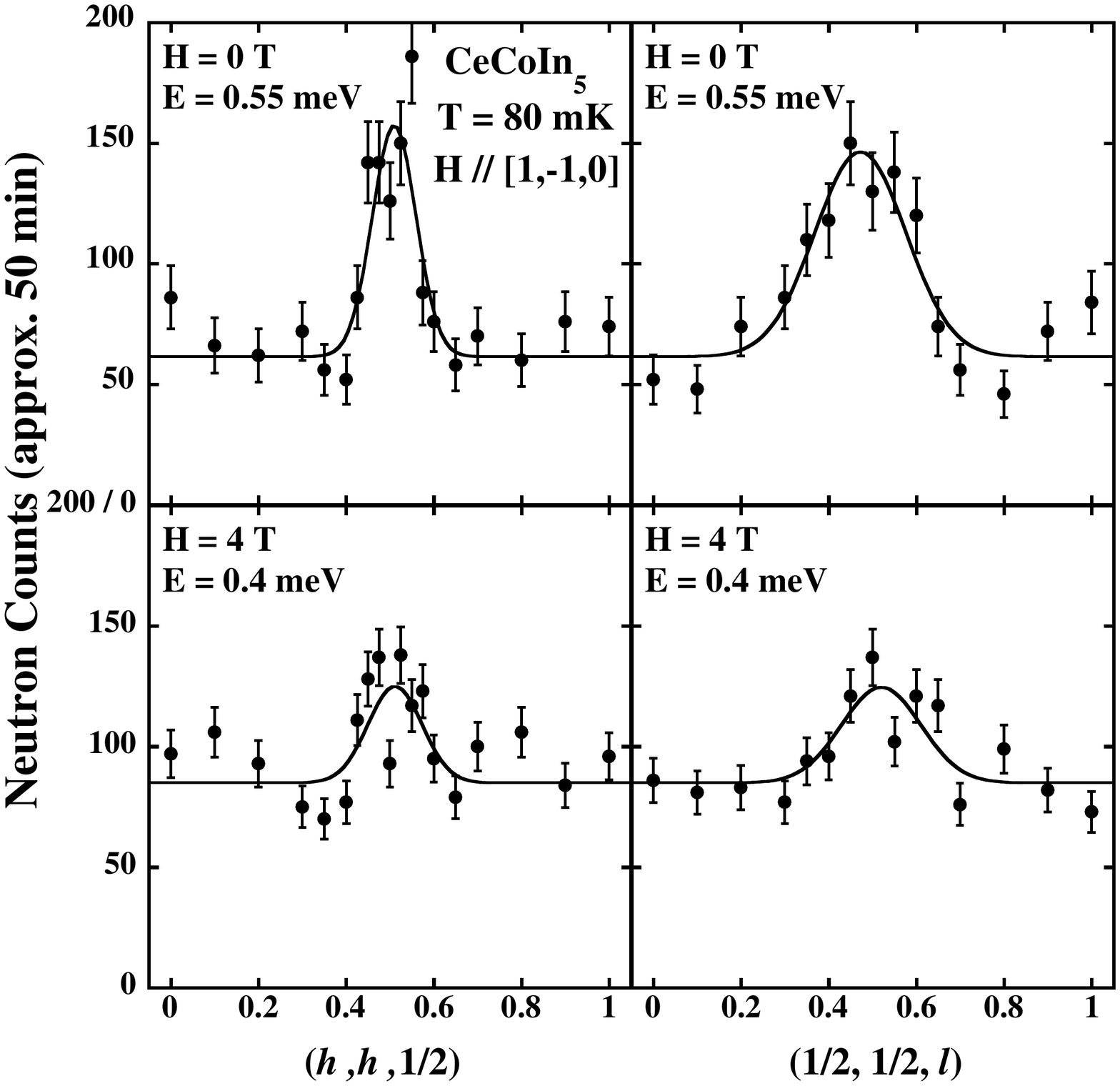}
\vspace{ -3.3cm}
\caption{Constant energy scans measured along the ($h$, $h$, 1/2) and (1/2, 1/2, $l$) lines at 0 T for $E$=0.55 meV (upper panels) and at 4 T for $E$=0.4 meV (lower panels). Lines are Gaussian fits.}
\end{figure}

The nature of the spin excitation in CeCoIn$_{5}$ is in debate. In the initial experimental \cite{Stock} and theoretical \cite{Eremin} works,  it is considered as a $S$=1 exciton below the particle-hole continuum in the line of the work performed on HTSC. However this picture is questioned for 3D systems \cite{Chubukov} and the excitation is proposed to be a magnon-type that appears below $T_{c}$ due to the reduction of Landau damping for energies below the superconducting gap. Given this situation, we will discuss on a phenomenological level the field dependence of the  three parameters, $\Delta$, $\Gamma$ and $\chi'$ extracted from our INS study.
\begin{figure}[t!]
\centering
\includegraphics[width=9cm]{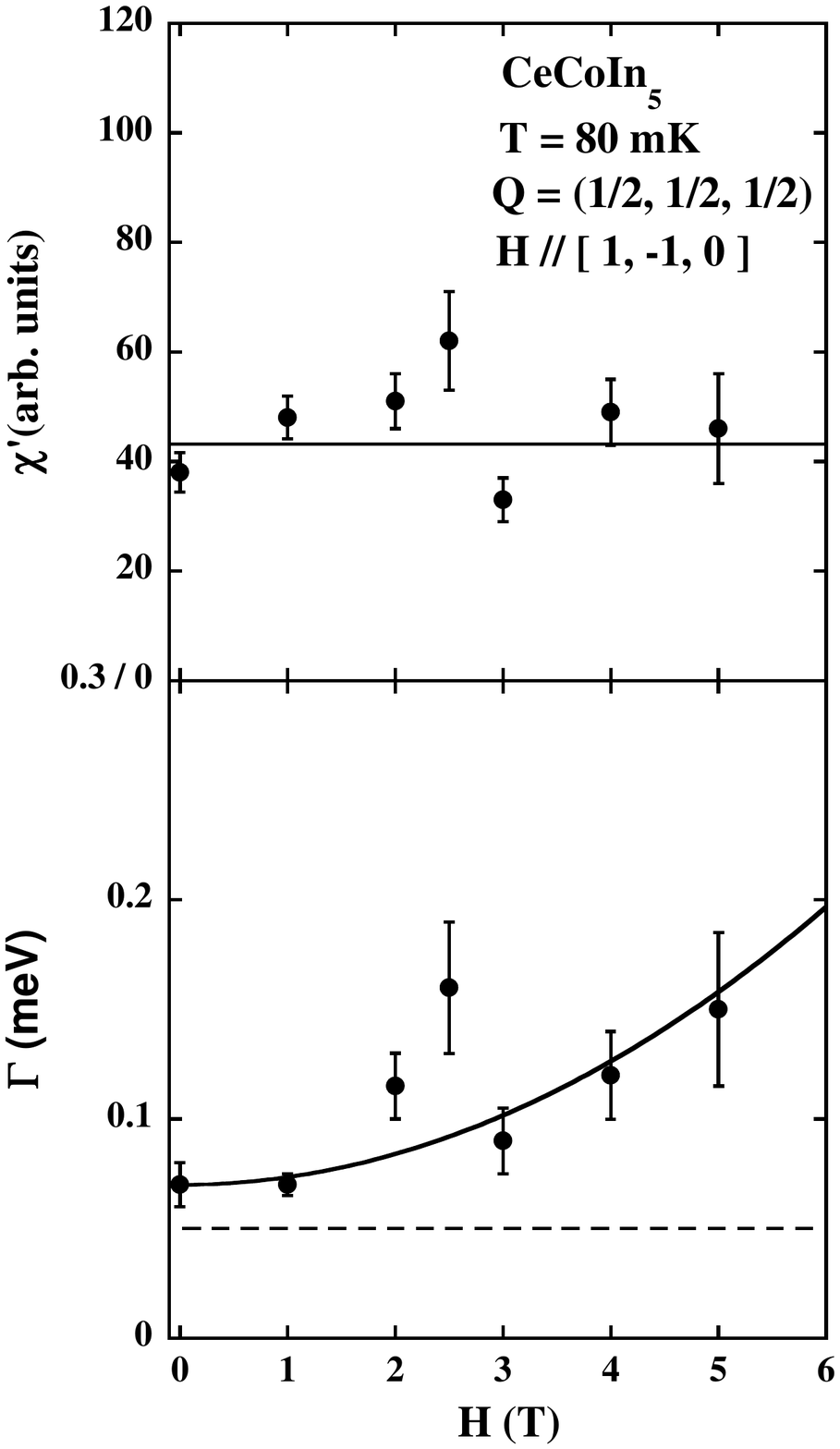}
\vspace{ -3.4cm}
\caption{Magnetic field dependence of $\chi'$ and $\Gamma$ measured at $\bf{Q}$=(1/2, 1/2, 1/2). Solid lines are guides for the eyes. The dashed line indicates the half width at half maximum of the incoherent signal.}
\end{figure}

\begin{figure}[t!]
\vspace{-2cm}
\centering
\includegraphics[width=7cm]{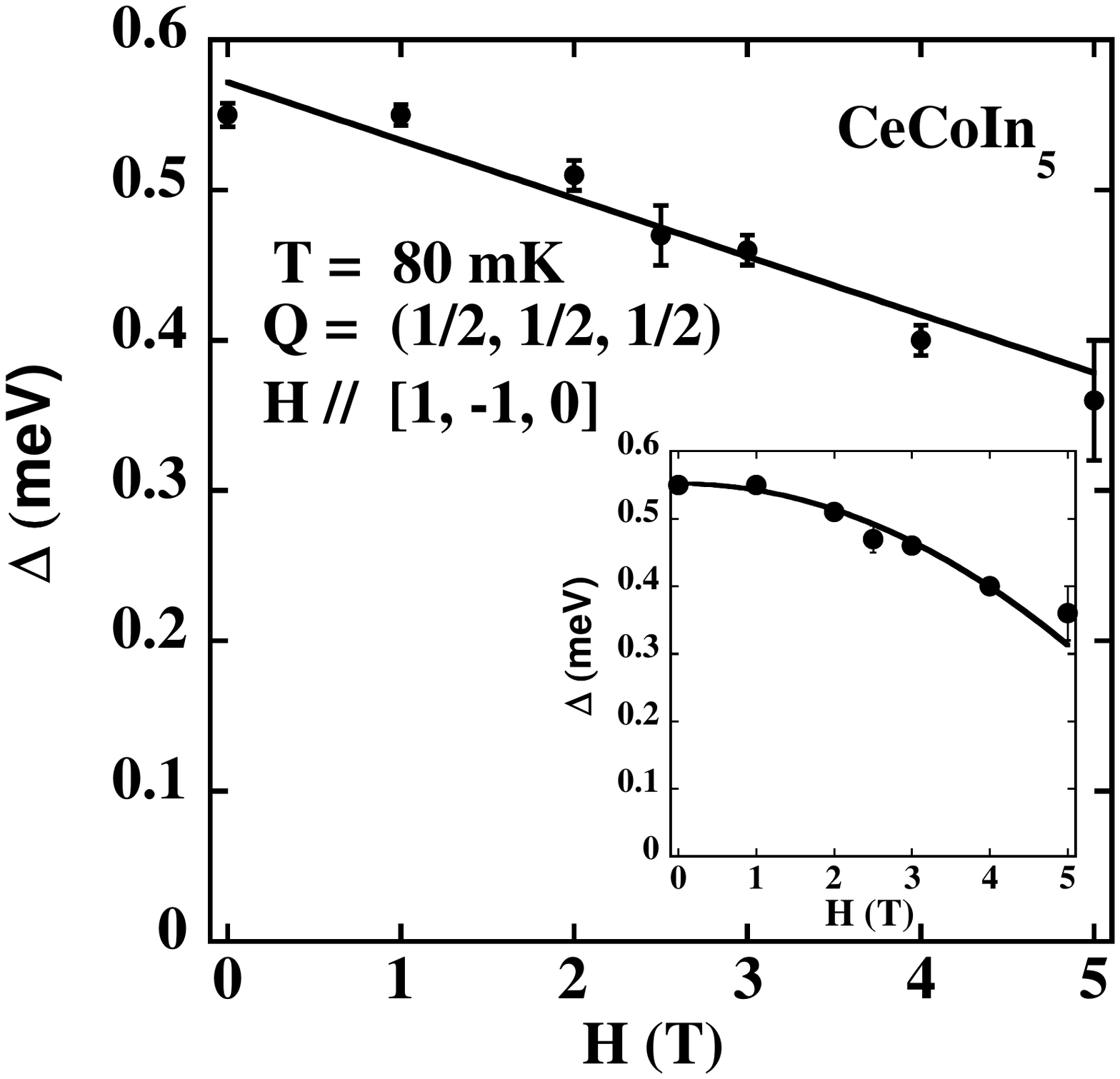}
\vspace{ -2cm}
\caption{Magnetic field dependence of the energy gap measured at $\bf{Q}$=(1/2, 1/2, 1/2). The line is a linear fit and the inset shows the same data with a quadratic fit.}
\end{figure}

The field dependence of the peak position can be either reproduced by a linear decrease, $\Delta(H)=\Delta(0)-\alpha H$, or by a quadratic decrease, $\Delta(H)=\Delta(0)-\beta H^2$. 
We obtained $\alpha$=0.039(3) meV/T and $\beta$=0.0096(6) meV/T$^2$. The extrapolation $\Delta(H)$=0 is obtained at 14.7 T in the linear fit and at 7.6 T in the quadratic fit. We recall that for a spin 1/2, the linear Zeeman shift is 0.058 meV/T for a gyromagnetic ratio $g$=2. Hence the rate of the linear decrease $\alpha$ is very similar to this value. Since the excitation is a magnetic mode, it is tempting to describe its field dependence by a Zeeman effect. However if the mode is intimately related to the superconducting gap, the main effect for its decrease would be orbital effects (at least for the magnetic fields up to 5 T applied in this experiment). This mechanism for the gap depression can be put forward since we do not observe any splitting of the mode, although its degeneracy is not know since its nature is not clear. In a simple view, the mode must collapse at 10.5 T on entering the LTHF. The only theoretical work calculating the field evolution of the spin gap that we are aware of concerns the study of the competition between superconductivity and spin density wave in a Ginzburg Landau approach\cite{Demler}. However this model, that does not reproduce our data, is probably not applicable to the case of CeCoIn$_{5}$ that lies in a range of parameters beyond the calculation. We will discuss further below the variation of $\Delta(H)$ obtained for other compounds, HF and HTSC.

The lineshape broadening is ascribed to the mixed state. Broadening of the spin resonance is theoretically predicted from the effect of the supercurrents around the vortices \cite{Eschrig}. Broadening is also observed in the Knight shift measurement at low fields (below the LTHF phase) and is well reproduced by a careful treatment of the vortex lattice \cite{Koutroulakis}. As concern the susceptibility $\chi'$, it is independent of the magnetic field. At 5 T, we have $\Delta$= 0.4 meV and $\Gamma$=0.12 meV and therefore the excitation is still well defined. In such a case the integrated intensity is proportional to $\chi'\times\Delta$ (for $T \ll E$, a condition realized here) and is therefore decreasing with increasing magnetic field up to 5 T. This means that the spectral weight is transfered to other part of the $(\textbf{Q}, E)$ space. For higher magnetic fields, first the signal will become quasielastic with $\Delta \approx \Gamma$ and also part of the intensity will go to the elastic peak at $\bf{Q}$=(0.44, 0.44, 1/2) in the LTHF phase. In this experiment the signal cannot be followed above 5 T due to large incoherent background inherent to the multi-crystal assembly and the use of focusing neutron optics. For higher magnetic field, competition between the superconducting ground state and the incommensurate magnetic order could lead to complex magnetic excitation spectra. We do not observed new modes for the spectra measured at $\bf{Q}$=(1/2, 1/2, 1/2) for $H$= 8 and 12 T at 80 mK in the range 0.4-1.2 meV (not shown in this paper).

Finally, we will compare our data with other strongly correlated electron systems for which magnetic field effect was studied on an excitation that exhibits signature of the superconducting state at low energy for zero magnetic field. For the unconventional HF superconductor PrOs$_{4}$Sb$_{12}$, the low energy crystal field excitation of energy 2.8$k_{B}T_{c}$ already exists in the normal phase and the suppression of the conduction electron scattering leads to a better definition of the peak in the superconducting phase \cite{Kuwahara}. Under magnetic field, the precise behavior of this excitation with respect to superconductivity is masked by the splitting of the triplet mode \cite{Raymond}. For UPd$_{2}$Al$_{3}$, the application of a magnetic field  parallel to the $b$-axis of the hexagonal structure, for which the critical field is 3.3 T, leads to a decrease of the spin gap, a broadening of the excitation and the appearance of quasielastic scattering \cite{Blackburn}. However it was shown that this latter scattering is maximum at 4.2 T and characterizes the fluctuations associated with the reorientation of the antiferromagnetic ordered moments. Below 2.08 T, this effect could be neglected and a precise field dependance of the gap is obtained. It bears some similarities with the one presented here for CeCoIn$_{5}$.  A linear fit of the gap would lead to $\alpha$=0.054(1) meV/T with a corresponding $\Delta$=0 extrapolated for $H$=6.6 T and a quadratic fit will give $\beta$=0.033(6) meV/T$^2$ with a corresponding extrapolated $\Delta$=0 for $H$=3.2 T. Note that while the values of $T_{c}$ (1.8 K in UPd$_{2}$Al$_{3}$, 2.3 K in CeCoIn$_{5}$) and the resonance modes are similar (0.35 meV in UPd$_{2}$Al$_{3}$, 0.6 meV in CeCoIn$_{5}$), the critical fields are much different for the two orientations compared here (3.3 T in UPd$_{2}$Al$_{3}$ and 11.6 T in CeCoIn$_{5}$). When comparing the two compounds, one must keep in mind that contrary to CeCoIn$_{5}$, UPd$_{2}$Al$_{3}$ is far from a magnetic instability. Finally for a field applied along the $c$-axis of UPd$_{2}$Al$_{3}$, there is no spin reorientation and no quasieleastic signal is observed under magnetic field, only the disappearing of the resonance is induced \cite{Hiess}.

In the HTSC superconductors of the YBa$_{2}$Cu$_{3}$O$_{6+x}$ familly, the application of a magnetic field on the resonance peak has no effect on the peak position \cite{Bourges,Dai}. This is understood given the large energy of the mode ($\approx$ 30 meV) compared to the magnitude of the applied magnetic field (approx. 10 T that corresponds to 0.58 meV for 1 $\mu_{B}$). However sizeable effect is observed on the width and intensity of the peak especially for field applied perpendicular to the CuO$_{2}$ planes. In the case of electron doped compounds Nd$_{2-x}$Ce$_{x}$CuO$_{4}$ and of the hole doped La$_{2-x}$Sr$_{x}$CuO$_{4}$, a spin gap less sharp than a resonance peak developps in the superconducting state. In Nd$_{1.85}$Ce$_{0.15}$CuO$_{4}$, it was found that the application of a magnetic field in the CuO$_{2}$ plane leads to a linear shift of the gap to zero for a field near $H_{c2}$ \cite{Motoyama}. In La$_{1.855}$Cr$_{0.145}$CuO$_{4}$, the suppression of the gap for field in the CuO$_{2}$ plane is more complex \cite{Chang} with a shape resembling the prediction of Demler et al.\cite{Demler}. Interestingly, when the spin gap collapses at $H_{c2}$, a static antiferromagnetic response appears at the same (incommensurate) wave-vector. It is quite tempting to draw a parallel with CeCoIn$_{5}$, where incommensurate order appears above 10.5 T. It is unfortunate that with the present setup, we cannot follow the dynamical response for higher magnetic field up to the softening of $\Delta$. It is worthwhile to note a very distinctive feature of  CeCoIn$_{5}$ : the field induced magnetic phase is sticked to the superconducting phase and disappears at $H_{c2}$. On the contrary in the related compound CeRhIn$_{5}$, the field induced antiferromagnetic phase that occurs above 2 GPa, starts well below $H_{c2}$ and extends well above $H_{c2}$ \cite{Park,Knebel}.

Our INS study of the field effect  on the spin resonance characteristic of the magnetic excitation spectrum of CeCoIn$_{5}$ in its superconducting phase superconducting evidences a gradual suppression of this mode. The proeminent features are the decrease of the energy gap and an increase of the lineshape broadening. Our data and in particular the field dependence of the energy gap contains microscopic information that deserves further  theoretical attention in view of better understanding the nature of the spin excitation. An experimental challenge will consist in studying how the resonance transforms into a static order in the LFHF phase and finding if new excitations characterize this phase.

We acknowledge discussions with P. Bourges, Y. Sidis, A. Hiess, V. Mineev, G. Knebel and J.-P. Brison. This work was supported by the ANR ECCE.


\begin{thebibliography}{9}
\bibitem{revue} For a review see e.g. J.L. Sarrao and J.D. Thompson, J. Phys. Soc. Japan \textbf{76}, 051013 (2007).
\bibitem{Stock} C. Stock et al., Phys. Rev. Lett. \textbf{100} (2008) 087001.
\bibitem{Metoki} N. Metoki et al., J. Phys. Soc. Japan \textbf{66}, (1997) 814.
\bibitem{Stockert} O. Stockert et al., Physica B \textbf{403} (2008) 973.
\bibitem{Sidis} Y. Sidis et al., C.R. Physique \textbf{8} (2007).
\bibitem{Ronnig} F. Ronnig et al., Phys. Rev. B \textbf{73} (2006) 064519 and references therein.
\bibitem{Revue2} For a review see Y. Matsuda and H. Shimahara, J. Phys. Soc. Japan \textbf{76} (2007) 051005.
\bibitem{Kenzelmann} M. Kenzelmann et al., Science \textbf{321} (2008) 1652.
\bibitem{Knebel2} G. Knebel et al., J. Phys. Soc. Japan \textbf{77} (2008) 114704. 
\bibitem{Eremin} I. Eremin et al., Phys. Rev. Lett. \textbf{101} (2008) 187001.
\bibitem{Chubukov} A. V. Chubukov and L.P. Gor'kov, Phys. Rev. Lett. \textbf{101} (2008) 147004.
\bibitem{Eschrig} M. Eschrig, M.R. Norman and B. Jank\`o, Phys. Rev. B \textbf{64} (2001) 134509.
\bibitem{Koutroulakis} G. Koutroulakis et al., Phys. Rev. Lett. \textbf{101} (2008) 047004.
\bibitem{Demler} E. Demler, S. Sachdev and Y. Zhang, Phys. Rev. Lett. \textbf{87} (2001) 067202. 
\bibitem{Kuwahara} K. Kuwahara et al.,Phys. Rev. Lett. \textbf{95} (2005) 107003.
\bibitem{Raymond} S. Raymond et al., J. Phys.: Condens. Matter \textbf{21} (2009) 215702.
\bibitem{Blackburn} E. Blackburn et al., Phys. Rev. B \textbf{74} (2006) 024406.
\bibitem{Hiess} A. Hiess et al., Phys. Rev. B \textbf{76} (2007) 132405.
\bibitem{Bourges} P. Bourges \textit{et al}, Physica B \textbf{234-236} (1997) 830-831.
\bibitem{Dai} P. Dai et al., Nature \textbf{406} (2000) 965.
\bibitem{Motoyama} E.M. Motoyama et al., Phys. Rev. Lett. \textbf{96} (2006) 137002.
\bibitem{Chang} J. Chang et al., Phys. Rev. Lett. \textbf{102} (2009) 177006.
\bibitem{Park} T. Park et al., Nature \textbf{440} (2006) 65.
\bibitem{Knebel} G. Knebel et al., Phys. Rev. B \textbf{74} (2006) 020501(R).

\end{thebibliography}
\end{document}